\documentclass[a4paper, 10pt, conference]{ieeeconf}      

\IEEEoverridecommandlockouts                              

\usepackage{amsmath} 
\usepackage{amssymb}  
\usepackage{graphicx}
\usepackage{array}
\usepackage{smartdiagram}
 \usepackage{booktabs} 
\usepackage{float} 
\usepackage{subfigure} 
\usepackage[hidelinks]{hyperref}
\usepackage[linesnumbered,ruled]{algorithm2e}

\title{\LARGE \bf
Dataset for predicting cybersickness from a virtual navigation task
}

\author{Yuyang Wang$^{1}$, Ruichen Li$^{2}$, Jean-Rémy Chardonnet$^{3}$ and Pan Hui$^{4}$
\thanks{*This work was not supported by any organization}
\thanks{$^{1}$Yuyang Wang, $^{2}$Ruichen Li and $^{4}$Pan Hui are with Hong Kong University of Science and Technology (Guangzhou), China
        {\tt\small yuyangwang@ust.hk,rli965@connect.hkust-gz.edu.cn, panhui@ust.hk}}%
\thanks{$^{3}$Jean-Rémy Chardonnet is with Arts et Métiers Institute of Technology, France
        {\tt\small jean-remy.chardonnet@ensam.eu}}%
}

\begin{document}

\maketitle
\thispagestyle{empty}
\pagestyle{empty}

\begin{abstract}

This work presents a dataset collected to predict cybersickness in virtual reality environments. The data was collected from navigation tasks in a virtual environment designed to induce cybersickness. The dataset consists of many data points collected from diverse participants, including physiological responses (EDA and Heart Rate) and self-reported cybersickness symptoms. The paper will provide a detailed description of the dataset, including the arranged navigation task, the data collection procedures, and the data format. The dataset will serve as a valuable resource for researchers to develop and evaluate predictive models for cybersickness and will facilitate more research in cybersickness mitigation.

\end{abstract}




\section{Introduction}

Cybersickness, also known as virtual reality sickness or simulator sickness, is a phenomenon where a person experiences symptoms similar to motion sickness after being exposed to virtual reality environments. The symptoms can include dizziness, headache, nausea, sweating, \textit{etc.}. Cybersickness occurs when the visual and vestibular systems in the brain receive conflicting information about self-motion, leading to a sensory conflict~\cite{reason1975motion}. This sensory conflict can cause discomfort and lead to the symptoms associated with cybersickness.

In recent years, there has been a growing interest in predicting cybersickness in virtual reality applications based on data mining technologies. Using datasets in this context allows researchers to analyze large amounts of data to identify patterns and correlations that may not be immediately apparent. This information can then advise the design of virtual reality environments that are less likely to induce cybersickness.

One area of research that has made effective use of datasets for predicting cybersickness is the use of artificial intelligence (AI) methods. AI algorithms can process and analyze large amounts of data to identify patterns and correlations that may not be immediately apparent. For example, in the context of cybersickness, AI methods can be used to analyze physiological responses, such as heart rate and skin conductance, to identify correlations with the presence and severity of cybersickness. one study that used AI methods for predicting cybersickness was conducted by Kim et al. (2019)~\cite{kim_deep_2019}. In this study, the authors used a deep learning algorithm to analyze data from the Simulator Sickness Questionnaire (SSQ)~\cite{kennedy1993simulator} and identify correlations between user behavior and the presence and severity of cybersickness. The results suggested that the deep learning algorithm could accurately predict the presence of cybersickness, with an accuracy rate of over 85\%.

The use of datasets for predicting cybersickness is necessary because it allows researchers to analyze and identify factors that contribute to the development of cybersickness. By collecting data on user behavior and physiological responses, researchers can determine the occurrence of cybersickness and the severity of its symptoms. This information can then guide the design of virtual reality environments and interaction techniques that can mitigate cybersickness. Additionally, data analysis can help researchers understand the underlying mechanisms of cybersickness, which can help develop more effective treatments and preventions.


\section{Dataset collection}

\begin{figure*}[htbp]
\centering
\subfigure[]{
\begin{minipage}[b]{0.45\linewidth}
\includegraphics[width=1\linewidth]{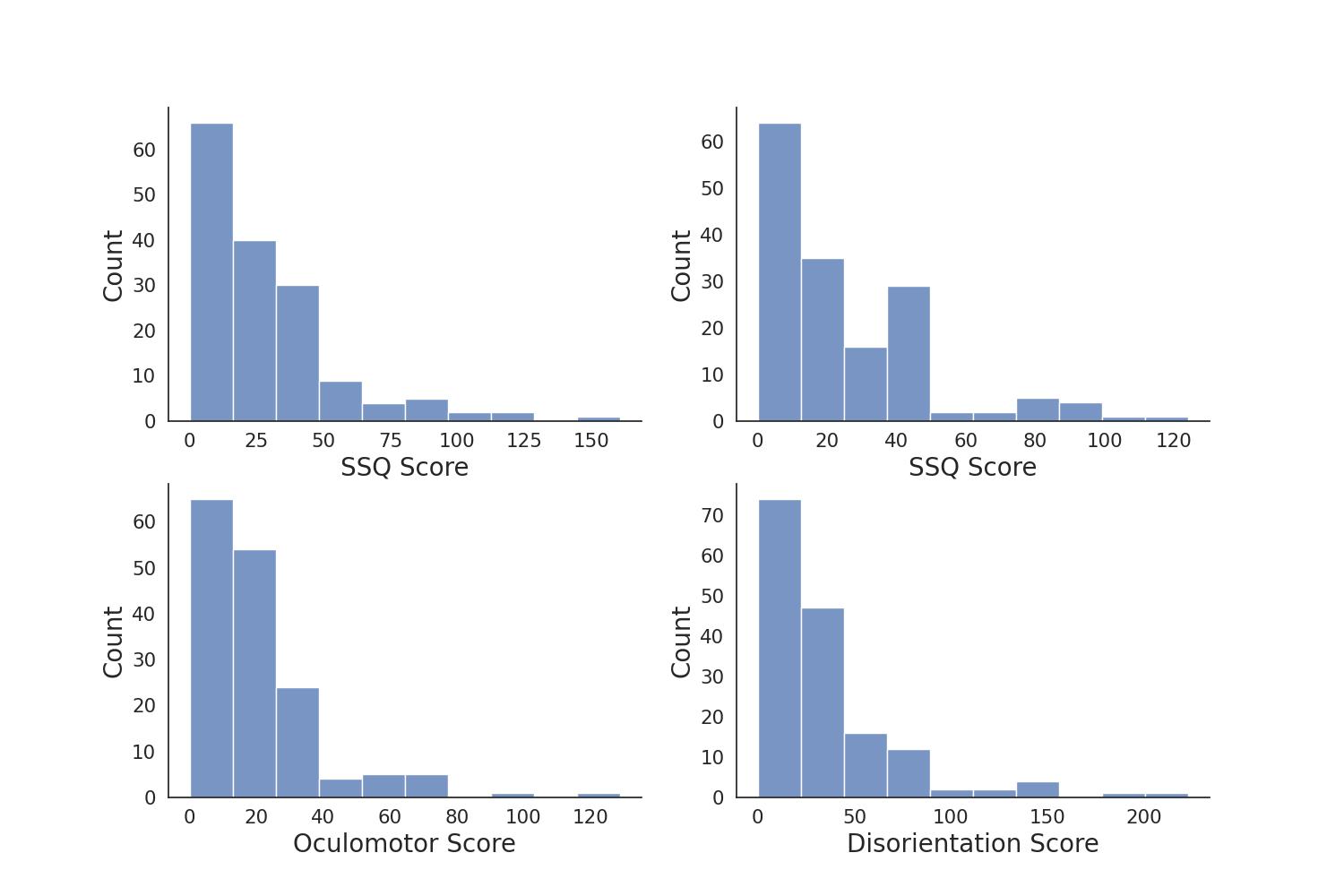}\vspace{4pt}
\end{minipage}}
\subfigure[]{
\begin{minipage}[b]{0.45\linewidth}
\includegraphics[width=1\linewidth]{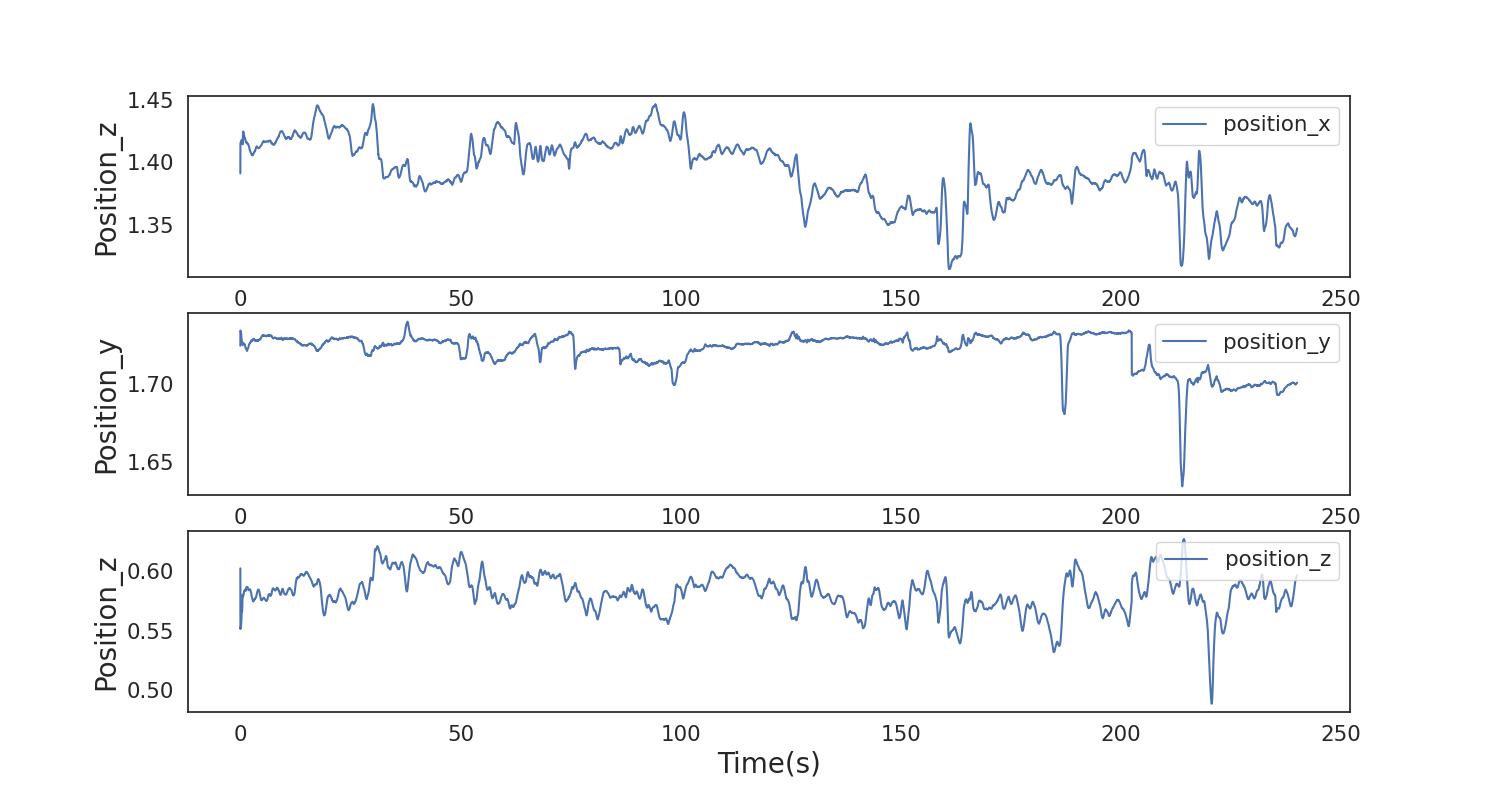}\vspace{4pt}
\end{minipage}}
\subfigure[]{
\begin{minipage}[b]{0.45\linewidth}
\includegraphics[width=1\linewidth]{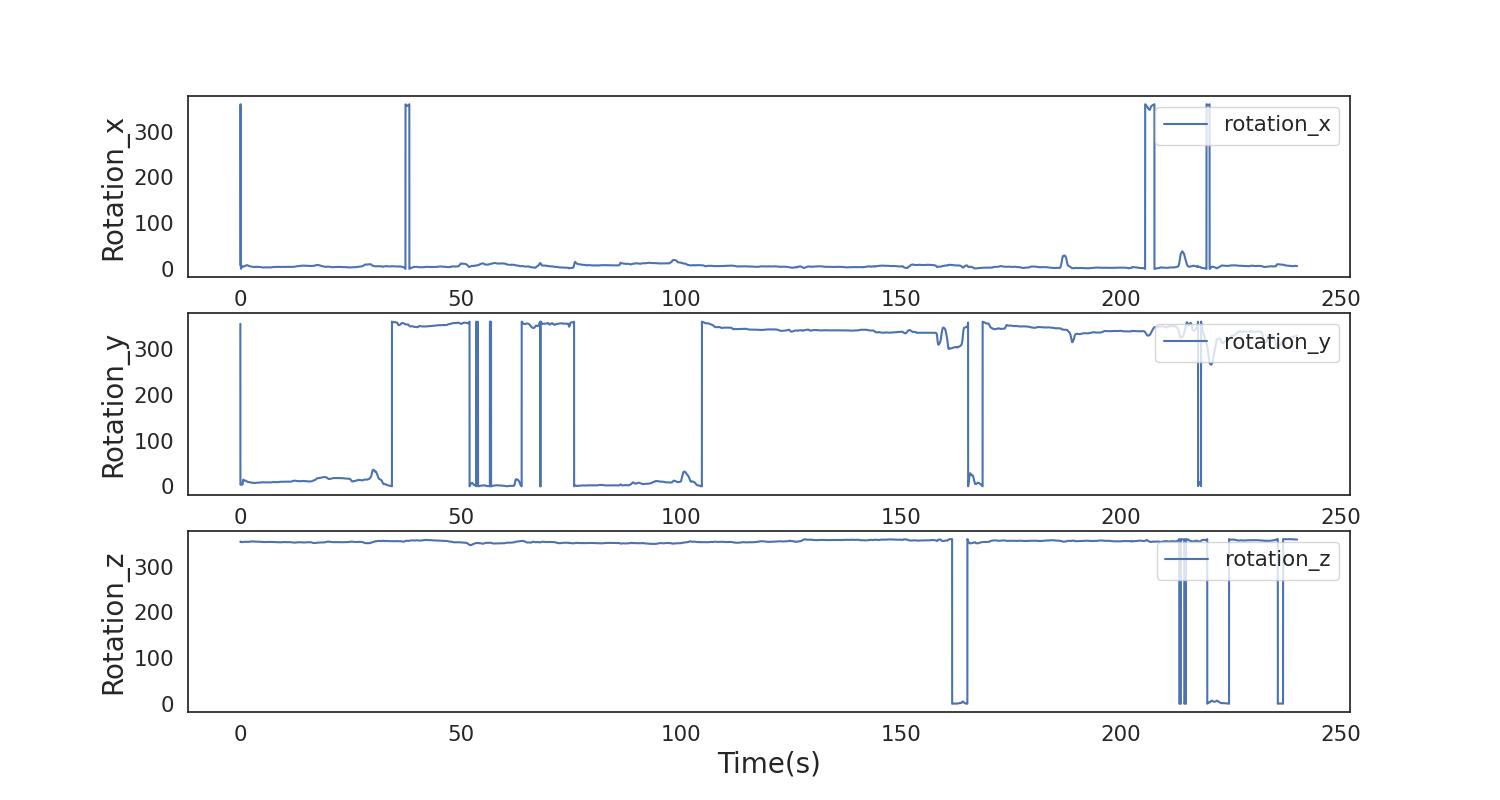}\vspace{4pt}
\end{minipage}}
\subfigure[]{
\begin{minipage}[b]{0.45\linewidth}
\includegraphics[width=1\linewidth]{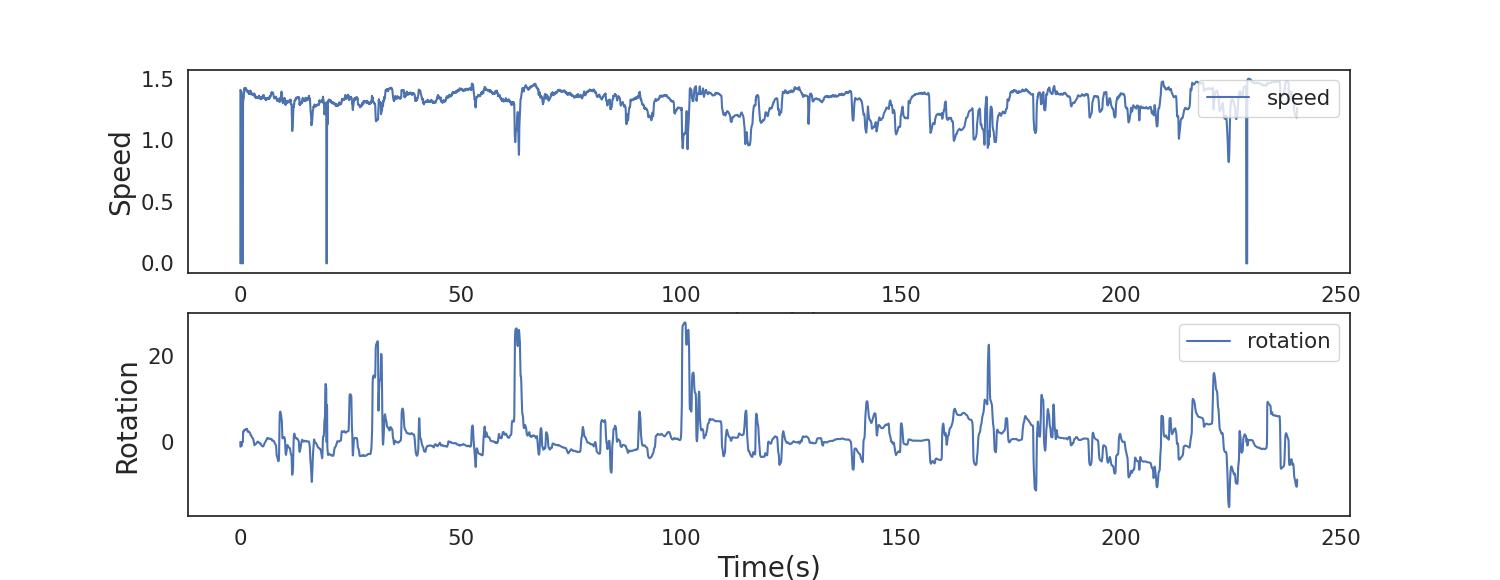}\vspace{4pt}
\end{minipage}}
\subfigure[]{
\begin{minipage}[b]{0.8\linewidth}
\includegraphics[width=1\linewidth]{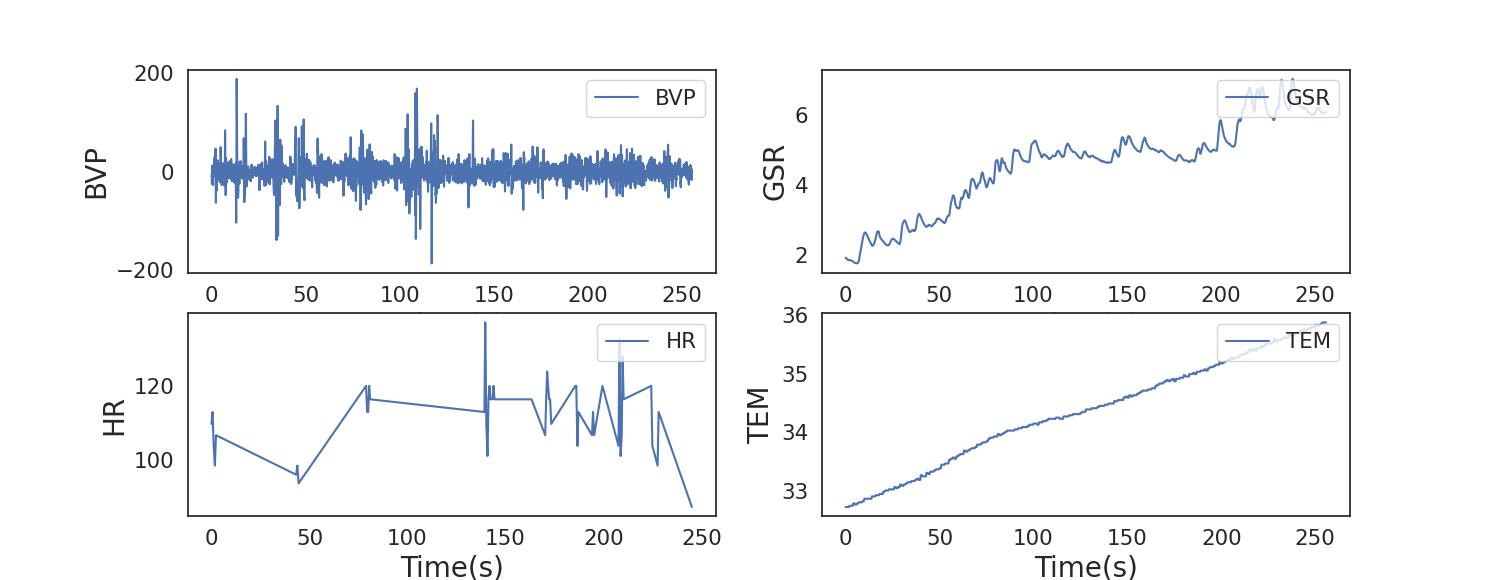}\vspace{4pt}
\end{minipage}}
\caption{Presentation of the data. (a) Score distribution of 159 samples. (b) Raw head position data of one sample. (c) Raw head rotation data of one sample. (d) Raw speed and raw rotation of one sample. (e) Biosignals of one sample.}\label{fig:onedata}
\end{figure*}

Fifty-three participants from the local city were recruited for a navigation task in a virtual reality environment. The participants had an average age of 26.3 years (SD = 3.3) with 26 of them being female. In order to gather more data, each participant was asked to participate three times over three separate days, resulting in a total of 159 samples. After completing the task, the participants were given various gifts as a token of appreciation. Before the experiment, they were required to fill out a pre-exposure questionnaire to assess their health status and experience with playing games and using VR devices. No health issues that could affect the experiment outcomes were reported through this questionnaire. All participants also signed a consent form prior to participating in the experiment.

\begin{figure*}[tb]
\centering
\includegraphics[width=0.9\textwidth]{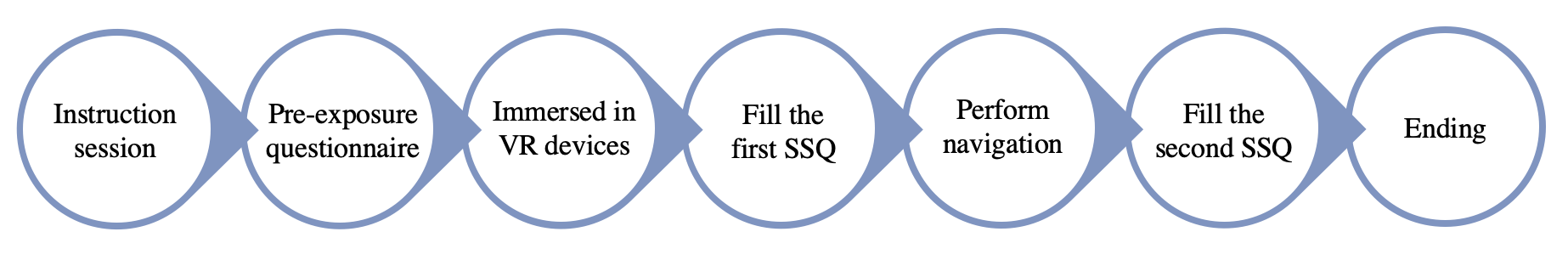}
\caption{Flowchart for the experiment process including user navigation with HTC Vive Pro and data collection with Empatica E4 wristband.}\label{fig:flowchart} %
\end{figure*}

\begin{figure}[h]
\centering
\includegraphics[width=0.9\columnwidth]{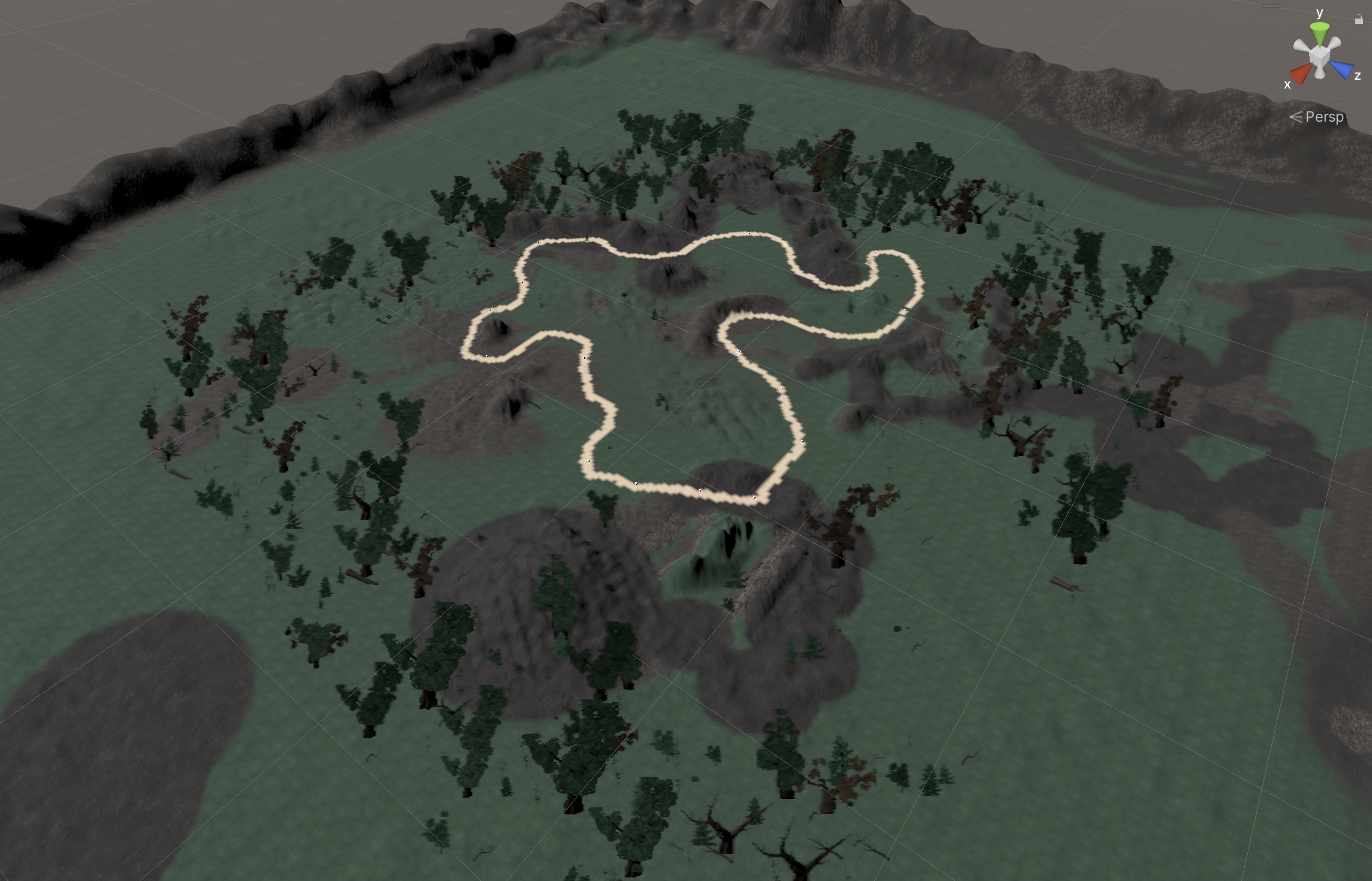}
\caption{Virtual scenario in which the participants navigate along the highlighted path.}
\label{fig:ve} %
\end{figure}

\begin{enumerate}
\item Prior to the test, the participants were given a brief explanation on how to manipulate navigation using the hand controllers of the HTC Vive Pro. In the event of cybersickness, they had the option to discontinue the experiment if they experienced discomfort or nausea.
\item The experimenter equipped the participants with an HTC Vive Pro headset and an Empatica E4 wristband for one participant. The Empatica E4 was capable of collecting Electrodermal Activity (EDA) data at a rate of $4$Hz, which was transmitted wirelessly to a computer for processing via Bluetooth.
\item The participants were then immersed in the virtual environment shown in \autoref{fig:ve} and began navigating along the trajectory indicated in brown. Navigation could be controlled in different directions through the touchpad on the HTC Vive Pro hand controller. Both the EDA signal and the longitudinal and rotational acceleration data were recorded simultaneously during navigation.
\item The navigation task lasted for four minutes and upon completion, the participants were removed from the head-mounted display.
\end{enumerate}

Additionally, the Simulator Sickness Questionnaire (SSQ) was used to evaluate the participants' symptoms of cybersickness by measuring three categories (nausea, oculomotor, and disorientation) through 16 questions. Participants filled out the SSQ both before and after the VR task to assess its psychological impact on them. The change in the score from the pre-exposure to the post-exposure was recorded and included in the data collection as the SSQ score:

\begin{equation}
    SSQ= SSQ_{post} - SSQ_{pre}
\end{equation}

\autoref{fig:flowchart} illustrates the general experimental process. 

\autoref{tab:content} presents the list of data collected from one experiment. The data collected can be categorized into four different types: \textbf{Head-Tracking}, \textbf{Locomation}, \textbf{Biosignal}, and \textbf{Cybersickness}. \textbf{Head-Tracking} includes raw and resampled head position, head rotation, and raw and resampled speed. \textbf{Locomation} consists of raw speed, resampled speed, raw rotation, and resampled rotation. \textbf{Biosignal} includes Galvanic Skin Response at a frequency of $4$Hz, Blood Volume Pulse at $64$Hz, Temperature at $4$Hz, and Heart Rate. \textbf{Cybersickness} includes the Simulator Sickness Questionnaire Score, Nausea Score, Oculomotor Score, and Disorientation Score. These data will provide valuable information to study the impact of immersive environments on physiological and psychological responses. \autoref{fig:onedata} depicts the SSQ distribution and the sequential data collected from one experiment. The full dataset is published a  \href{https://github.com/coreturn/Cybersickness_Dataset}{github repository}\footnote{\href{https://github.com/coreturn/Cybersickness_Dataset}{https://github.com/coreturn/Cybersickness$\_$Dataset}}, which has been used in two publications for predicting cybersickness with different inputs~\cite{wang_using_2021,hadadi_prediction_2022}.

\begin{table}[htbp]
\centering
\caption{List of data content collected from one experiment}\label{tab:content}
\renewcommand\arraystretch{1.3}
\scalebox{0.9}{
\begin{tabular}{cl} 
\toprule
\textbf{Data Type}          & \multicolumn{1}{c}{\textbf{Data}}            \\ 
\hline
\textbf{Head-Tracking Data} & \begin{tabular}{l}Raw Head Position~(i.e., x, y and z)\\Resampled Head Position~(i.e., x, y and z)\\Head Rotation~(i.e., x, y and z)\\Resampled Head Rotation~(i.e., x, y and z)\end{tabular}   \\ 
\hline
\textbf{Motion Data}        & \begin{tabular}{l}Raw Speed\\Resampled Speed\\Raw Rotation\\Resampled Rotation\end{tabular}   \\ 
\hline
\textbf{Biosignal Data}     & \begin{tabular}{l}Galvanic Skin Response~(4Hz)\\Blood Volume Pulse~(64Hz)\\Temperature~(4Hz)\\Heart Rate\end{tabular}   \\ 
\hline
\textbf{Cybersickness Data} & \begin{tabular}{l}Simulator Sickness Questionnaire Score\\Nausea Score\\Oculomotor Score\\Disorientation Score\end{tabular}      \\
\bottomrule
\end{tabular}
}
\end{table}

\section{Dataset Summary and usage}

The use of datasets for predicting cybersickness has been a valuable tool for researchers in the virtual reality community. By analyzing large amounts of data, researchers have identified correlations between physiological responses and the presence and severity of cybersickness. This work presents preliminary results from the dataset analysis, which suggest that the dataset is a rich and valuable resource for predicting cybersickness.

\addtolength{\textheight}{-12cm}   







\bibliographystyle{IEEEtran}
\bibliography{reference}

\end{document}